\documentclass[onefignum,onetabnum]{siamart190516}

\usepackage[caption=false]{subfig}

\usepackage{lipsum}
\usepackage{amsfonts}
\usepackage{graphicx}
\usepackage{epstopdf}
\usepackage{algorithmic}
\usepackage{xspace}

\ifpdf
  \DeclareGraphicsExtensions{.eps,.pdf,.png,.jpg}
\else
  \DeclareGraphicsExtensions{.eps}
\fi

\usepackage{enumitem}
\setlist[enumerate]{leftmargin=.5in}
\setlist[itemize]{leftmargin=.5in}

\newsiamremark{remark}{Remark}
\newsiamremark{hypothesis}{Hypothesis}
\crefname{hypothesis}{Hypothesis}{Hypotheses}
\newsiamthm{claim}{Claim}

\headers{Clicker Question Bank for Numerical
  Analysis}{M. A. Hossain, P. M. Menz and J. M. Stockie}

\title{An Open-Access Clicker Question Bank for \\
  Numerical Analysis\thanks{Submitted to the editors May 24, 2020.
    \funding{This work was funded by an Open Educational Resource (OER)
      Grant from Simon Fraser University and a Discovery Grant from the
      Natural Sciences and Engineering Research Council of Canada
      (NSERC).}}}

\author{M. Alamgir Hossain\thanks{Department of Mathematics, Simon Fraser
    University, Burnaby, BC, Canada
  (\email{mahossai@sfu.ca}, \email{pmenz@sfu.ca}, \email{jstockie@sfu.ca}).}
\and Petra M. Menz\footnotemark[2]
\and John M. Stockie\footnotemark[2]}

\usepackage{amsopn}

\newcommand{\courseID}{MACM~316\xspace} 
\newcommand{\coursename}{Numerical Analysis~I\xspace}

\ifpdf
\hypersetup{
  pdftitle={An Open-Access Clicker Question Bank for Numerical
    Analysis}, 
  pdfauthor={M. A. Hossain, P. M. Menz and J. M. Stockie}
}
\fi

\begin{document}

\maketitle

\begin{abstract}
  We present a question bank consisting of over 250 multiple-choice and
  true-false questions covering a broad range of material typically
  taught in an introductory undergraduate course in numerical analysis
  or scientific computing. The questions are ideal for polling students
  during lectures by means of a student response system that uses
  \emph{clicker} remotes or smartphones running a suitable app. We
  describe our experiences implementing these clicker questions in a
  recent class and provide evidence of their effectiveness in terms of
  testing students' prior knowledge, gauging understanding of new
  material, increasing participation, and especially improving student
  satisfaction. Our conclusions are supported by a mid-semester student
  survey as well as anecdotal observations. The question bank has been
  released as an open-access educational resource under a Creative
  Commons
  license (BY-NC-SA) for free use by the mathematics community.
\end{abstract}



\newcommand{\mysubsection}[1]{\subsection*{#1}}
\renewcommand{\mysubsection}[1]{\subsection{#1}}

\section{Introduction}
\label{sec:intro}

Student response systems, commonly referred to as \emph{clickers}, have
become commonplace in university classrooms as a way to interactively
probe students' understanding of course material and to engage them more
actively in lectures. These systems allow polling of students through a
combination of specialized hardware and/or software that record,
aggregate and display student responses in real-time.  Clickers serve a
variety of useful pedagogical purposes such as: testing prior knowledge,
gauging student understanding of new concepts, providing instantaneous
and visual feedback to students, increasing classroom participation,
etc. A major advantage of being able to poll the entire class in this
manner is that one can immediately identify situations where a
significant number of students are confused or lacking understanding on
an important point, thereby empowering the instructor to adjust their
lecture on-the-spot to address the problem.

Although clickers have been utilized extensively and effectively in many
university classes, it has been predominantly in lower division courses;
and in the case of mathematics, attention has been predominantly on
development of clicker questions for introductory courses in calculus
and linear algebra. There is a particularly conspicuous gap in
openly-available clicker resources that cover material in the area of
numerical analysis, which is identified next.

\mysubsection{Review of Clicker Resources: The Numerical Analysis Gap}

The use of short multiple-choice or true-false questions in a lecture
setting was first proposed for physics courses by Eric Mazur in the form
of \emph{ConcepTests} \cite{crouch-mazur-2001, mazur-1996}. This
approach was rapidly incorporated in other disciplines, and the polling
of students during lectures was made easier and more interactive with
the development of electronic student response systems. These clicker
devices, along with accompanying software, allow students' responses to
be compiled immediately, presented as a summary table or bar chart, and
then discussed further in class. An excellent literature review of
clickers and their pedagogical uses in a wide range of disciplines is
provided by Caldwell~\cite{caldwell-2007}.  Clicker questions have
become pervasive in mathematics instruction and there are numerous
studies of clickers in entry-level undergraduate mathematics courses
\cite{bode-etal-2009, lucas-2009, menz-jungic-wiebe-2009,
  miller-etal-2006, roth-2012, schlatter-2002}. These studies have
established that use of clicker polling in lectures is ``yet another
source of a valuable learning tool for
students''~\cite{menz-jungic-wiebe-2009} and can generate a number of
positive outcomes for both instructors and students. The most obvious
and prosaic benefit for instructors is in terms of increased attendance,
although they also clearly benefit from gaining an immediate measure of
students' current understanding of course concepts.  From the student's
point of view, the advantages include immediate and regular feedback on
their learning, an ability to gauge their own knowledge against that of
their peers, and improved satisfaction with their learning
experience. Regular participation in clicker polls can also yield the
interesting spin-off effect of stimulating other forms of in-class
participation such as peer discussions or think-pair-share, thereby
engaging students even more actively in lectures.

Devising interesting and insightful clicker questions can pose
significant challenges. A concise and practical resource that provides
useful tips on how best to formulate questions and implement them
effectively in lectures is the \emph{``Clicker Resource Guide''} by
Wieman et al.~\cite{cwsei-clicker-guide-2017}.  In this guide, the
authors identify ten {question types} that serve various purposes in a
classroom setting, such as:
\begin{enumerate}
\item testing comprehension of assigned readings, 
\item assessing recall of lecture points,
\item performing a step in a calculation, 
\item surveying student opinions,
\item revealing pre-existing knowledge or thinking, 
\item testing conceptual understanding, 
\item applying ideas in a new context, 
\item predicting results of a demo or simulation, 
\item drawing on knowledge from everyday life, 
\item relating different representations (mathematical, graphical,
  algorithmic, etc.).
\end{enumerate}
Devising questions that satisfy such a wide variety of purposes for all
topics in a course can be a daunting task, but thankfully there already
exist extensive collections of openly available questions for certain
subjects.  The two most notable resources in mathematics are the
GoodQuestions project for calculus \cite{terrell-conelly-goodq}, and the
MathQUEST/MathVote \cite{cline-zullo-2011, zullo-cline-mathquest}
project which covers a broader range of the first- and second-year
undergraduate curriculum in calculus, linear algebra, differential
equations and statistics.  Several textbooks come bundled with a clicker
resource~\cite{anton-bivens-davis-2012, hughes-hallett-etal-2018,
  rogawski-adams-2015} but these questions are subject to copyright and
so are not freely available.

When it comes to upper-year undergraduate courses like numerical
analysis, there are comparatively few open collections of clicker-type
questions. Some multiple-choice and true-false questions can be found
among the exercises in certain numerical analysis textbooks, the most
noteworthy being Heath's text \cite{
  heath-2018}. The only freely-available question bank we are aware of
is distributed along with an open textbook by Kaw et
al.~\cite{kaw-holistic}, but both the number of questions and diversity
of topics are somewhat limited.

\mysubsection{Aims and Outline}

The main purpose of our paper is to remedy this ``numerical analysis
gap'' and our aims in this respect are three-fold. First of all, we
provide an account in \cref{sec:course,sec:qbank} of our efforts to
construct a question bank that covers much of the standard curriculum in
a third-year undergraduate course at Simon Fraser University called
\emph{\coursename} (\courseID for short). Questions span the theory and
application of algorithms for scientific computing and are posed as
either true-false or multiple-choice questions (with 3 to 5
choices). Our intent is to come up with a sufficiently large collection of
questions (at least 250 in number) that not only provides comprehensive
coverage of the course material but also permits the instructor some
flexibility in terms of both level of difficulty and question
type. Because algorithms and Matlab programming are an integral part of
\courseID, we also include questions that test students' understanding
of basic algorithms in the form of Matlab code, such asking them to identify
bugs in erroneous code or to interpret tabulated or graphical output.

Secondly, we are releasing the clicker question bank as an open
educational resource (OER) that should be of significant value to other
teachers of numerical analysis from the applied mathematics
community. We describe in \cref{sec:qbank} how the questions and
solutions are conveniently written in \LaTeX, and macro files and slide
templates are provided so that questions can be easily incorporated into
other instructors' lectures. \Cref{sec:polling} is a brief ``how-to
guide'' that explains how we have integrated the clicker questions into
our own lectures, along with some helpful in-class tips.  By releasing
this resource under a Creative Commons Share-Alike (CC BY-NC-SA)
license, we are strongly encouraging readers to contribute to this OER
themselves by making corrections or improvements to existing questions,
and more importantly by proposing new questions of their own that expand
on the topics already covered. The following web site provides the
complete resource in both \verb+zip+ and \verb+tar.gz+ formats for easy
download:

\begin{center}
  \url{http://www.sfu.ca/~jstockie/NAclicker}
\end{center}

Our third and final aim is to share our experiences implementing the
clicker question bank in a numerical analysis class during a recent
semester (January to April 2020) and to illustrate how it has led to
significant improvements in student satisfaction as well as the
classroom learning environment. Support for these claims is provided in
\cref{sec:survey}, which summarizes the results of a mid-semester
student survey and also provides anecdotal observations comparing our
experience teaching with clickers to several previous
offerings of the same course without clickers.

If you have ever been tempted to experiment with clickers in your own
numerical analysis course but were intimidated by the amount of work
involved in writing good questions, then this just might be the OER for
you! 

\section{Course Overview}
\label{sec:course}

We begin with a description of SFU's numerical analysis course \courseID
to set the context for the development of the clicker question bank.
\courseID is a course called \emph{\coursename} that covers the
elementary topics from numerical analysis listed in the syllabus
provided in \cref{tab:syllabus}.
\begin{table}[tbhp]
  \centering
  \label{tab:syllabus}
  \caption{Syllabus for \courseID. The numbering of topics is the same as
    that used in the question bank, and the corresponding section number
    is listed from Burden, Faires and Burden~\cite{burden-faires-2015}.} 
  \begin{tabular}{|lllc|}\hline
    \#  & \multicolumn{2}{l}{Topic} & BFB Section\\\hline
    1. & \multicolumn{3}{l|}{Introduction:}\\
    & 1a. & Floating-point arithmetic and errors & 1.2\\
    & 1b. & Review of concepts from calculus & 1.1\\
    \hline
    2. & \multicolumn{3}{l|}{Nonlinear equations:}\\
    & 2a. & Bisection method & 2.1 \\
    & 2b. & Fixed point method & 2.2 \\
    & 2c. & Secant and Newton's methods & 2.3 \\
    & 2d. & Convergence of iterative algorithms & 1.3\\
    \hline
    3. & \multicolumn{3}{l|}{Systems of linear equations:}\\
    & 3a. & Review of linear algebra & 6.1, 7.1\\
    & 3b. & Gaussian elimination, pivoting and matrix factorization &
    6.1, 6.2, 6.5\\
    & 3c. & Plotting power laws (interlude) & \\
    & 3d. & Iterative methods for sparse systems & 7.3, 7.4\\
    \hline
    4. & \multicolumn{3}{l|}{Function approximation:}\\
    & 4a. & Polynomial (smooth) interpolation & 3.1, 3.3\\
    & 4b. & Spline (piecewise) interpolation & 3.5 \\
    & 4c. & Least squares fitting & 8.1, 8.2\\
    \hline
    5. & \multicolumn{3}{l|}{Differentiation and integration:}\\
    & 5a. & Numerical differentiation & 4.1, 4.2\\
    & 5b. & Numerical integration or quadrature & 4.3, 4.4, 4.6\\
    \hline
    6. & \multicolumn{3}{l|}{Ordinary differential equations (ODEs):}\\
    & 6a. & Background on ODE initial value problems & 5.1\\
    & 6b. & Euler's method & 5.2\\
    & 6c. & Taylor and Runge-Kutta methods & 5.3, 5.4\\
    & 6d. & Systems of first-order ODEs & 5.9\\
    \hline
  \end{tabular}
\end{table}
These are fairly standard topics in many introductory numerical analysis
courses taught in universities around the world, and the topics map
easily to the contents of most textbooks, one example being the seminal
\emph{``Numerical Analysis''} by Burden, Faires and Burden
\cite{burden-faires-2015} (which is the required text for
\courseID). The ordering of topics in the course syllabus is slightly
different from that in many books including \cite{burden-faires-2015}
and is solely a personal choice of the instructor (JMS).  A teaching
semester at SFU consists of roughly 40 hours of instruction spread over
13 weeks and courses are parceled into 50-minute lectures taught three
times per week.

\mysubsection{Student Demographics} 

\courseID is a core course in the \emph{Mathematics and Computing
  Science} (MACM) program at Simon Fraser University. Roughly two-thirds
of students are enrolled in the School of Computing Science, where the
course is a graduation requirement. The remaining students are divided
almost equally between programs in the Faculties of Science and Applied
Science (Engineering). The prerequisites for \courseID are two
first-year classes in differential and integral calculus, plus an
introductory linear algebra class.  Some computing experience is
required (or at least strongly recommended) because algorithms are
central and students are obliged to submit Matlab code as part of their
assignment solutions.  Major theorems and results on error estimates and
convergence are all covered, but the emphasis is more on practical
implementations of algorithms than on the theory behind them.
 

The usual enrollment in \courseID hovers around 200 students, with
roughly 25\%\ in year 2 and the remainder in years 3-4 of their
undergraduate programs. The main point of this remark is that the
class is a relatively large one, and it is also an upper-year course so that
most students are already ``more experienced''.  This distinguishes the
student cohort in \courseID from that in most other clicker-based
courses because clickers are most often implemented in large
introductory classes~\cite{beckert-fauth-olsen-2009, bode-etal-2009}.

\mysubsection{Grading Scheme} 

The serious computing component of this class is reflected in the
grading scheme in \cref{tab:grading} which puts significant weight on
computing assignments.
\begin{table}[bthp]
  \centering
  \caption{Grading scheme for \courseID.}
  \label{tab:grading}
  \begin{tabular}{|lr|}\hline
    Quizzes (weekly, 10 mins long, in class) & 18\%\\
    Computing assignments (bi-weekly) & 24\%\\
    Clicker questions (participation only) & 3\%\\
    Midterm test & 25\%\\
    Final exam & 40\%\\\hline
  \end{tabular}
\end{table}
We also highlight the 3\%\ of the final grade that is assigned to
participation in the in-class clicker polls. These 3 points are an
``all-or-nothing'' grade that a student receives only if they respond to
at least 75\%\ of clicker questions posed during semester. Our aim is to
encourage students to be present in class and to participate in the
clicker polls, but to otherwise avoid imposing undue pressure on
students to obtain the correct answer.

\section{Structure of the Clicker Question Bank}
\label{sec:qbank}

The clicker question bank consists of a mix of both mul\-ti\-ple-choice
and true-false questions based on the material from the syllabus in
\cref{tab:syllabus}, which lists topics that are easily mapped onto
chapters/sections from most numerical analysis textbooks.  In authoring
the questions, we aimed to employ a variety of the question types
identified in the introduction (numbered 1--10) although we concede that
this initial release of the question bank includes very few of types 1,
4 or~9.

Approximately 80\%\ of questions were authored by ourselves, with the
remaining 20\%\ derived from a few specific sources. The bulk of these
latter questions were adapted from the numerical analysis texts by
Heath~\cite{heath-2018} and Kaw et al.~\cite{kaw-holistic}.  A smaller
number of questions on background material from calculus and linear
algebra were based on material from the
GoodQuestions~\cite{terrell-conelly-goodq} and MathQUEST/MathVote
projects~\cite{zullo-cline-mathquest}.

\begin{figure}[bthp]
  \centering
  \fbox{\includegraphics[width=0.96\textwidth]{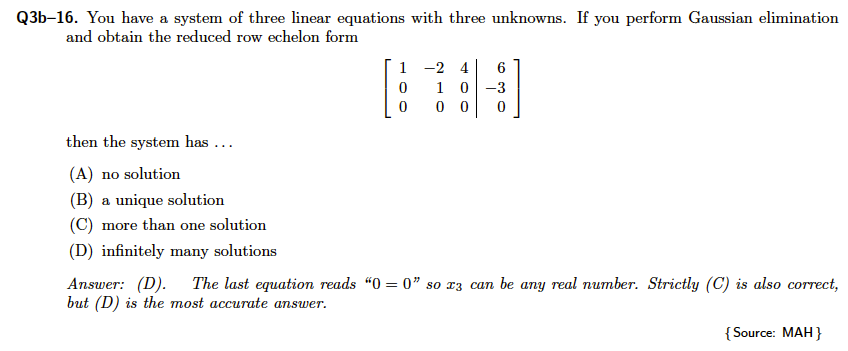}}
  \caption{Formatted output from {\tt clickerQbank.pdf} for a sample
    multiple-choice clicker question having 4 choices, typeset using the
    macro {\tt\textbackslash{}qitemMCfour}.}
  \label{fig:qexample1}
\end{figure}

\begin{figure}[bthp]
  \centering
  \subfloat[First slide with question only.]{%
    \begin{minipage}{0.5\textwidth}
      \includegraphics[width=\textwidth]{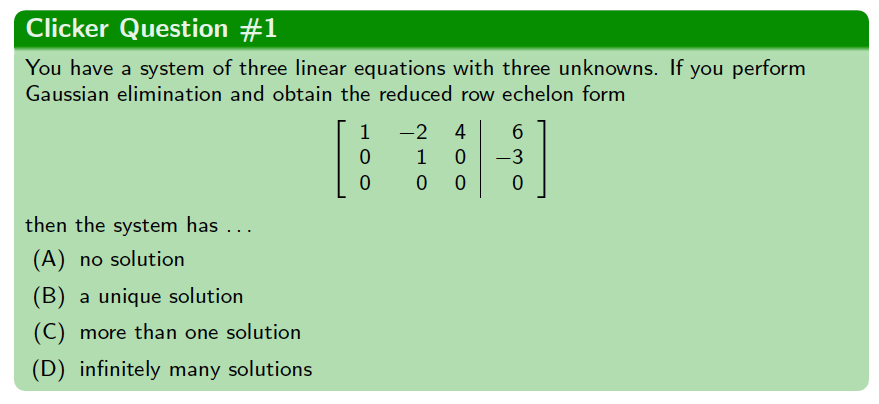}\\
      \rule{0cm}{0.9cm}
    \end{minipage}%
    \label{fig:qexample2a}} 
  \subfloat[Second slide showing question and answer.]{%
    \begin{minipage}{0.5\textwidth}
      \includegraphics[width=\textwidth]{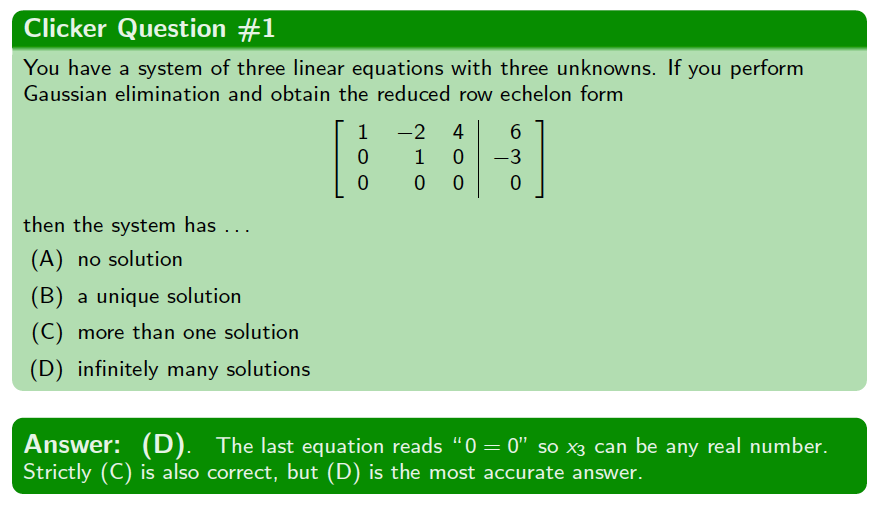}  
    \end{minipage}%
    \label{fig:qexample2b}} 
  \caption{Beamer slides from {\tt clickerbeamer.pdf} for the
    same multiple-choice question in
    {\protect\cref{fig:qexample1}}.}
  \label{fig:qexample2}
\end{figure}

The clicker question bank (consisting of \LaTeX\ source, image files and
Matlab code) can be obtained from the URL provided at the end of the
introduction.  Below are a few main features of the question bank that
will help readers who are interested in implementing the questions in
their own courses and/or adding questions of their own:
\begin{itemize}
\item In total there are 267 questions, of which 242 are multiple-choice
  and the remainder are true-false. The multiple-choice questions allow
  for 3, 4 or 5 choices (labelled A,B,C,D,E).

\item Questions are encoded in \LaTeX\ to ensure high-quality
  typesetting of the mathematics and to allow them to be easily
  modified and cut-and-paste into other \LaTeX\ documents.  Several
  \LaTeX\ files deserve explanation:
  \begin{enumerate}[label=(\alph*)]
  \item \verb+clickerQbank.{pdf,tex}+: A document containing all
    clicker questions that generates a ``compact'' 90-page listing of
    questions and solutions.  Each question is defined as an item within
    a list environment called \verb+clicklist+, and there is a macro
    named \verb+\qitem...+ defined for each question type. For example,
    a multiple-choice question with four choices (A,B,C,D) is encoded
    using the macro \verb+\qitemMCfour+ and is passed 8 arguments as
    follows:

    \vspace*{0.2cm}
\hrulefill
\begin{verbatim}
\begin{clicklist}
  ...
\qitemMCfour{You have a system of three linear equations
   with three unknowns. If you perform Gaussian 
   elimination and obtain the reduced row echelon form
   \begin{gather*}
     \left[ \begin{array}{rrr|r}
         1 & -2 & 4 &  6 \\
         0 &  1 & 0 & -3 \\
         0 &  0 & 0 &  0  
       \end{array} \right]      
   \end{gather*}
   then the system has \dots}% ARG1: question statement
  {no solution}               % ARG2: response (A)
  {a unique solution}         % ARG3: response (B)
  {more than one solution}    % ARG4: response (C)
  {infinitely many solutions} % ARG5: response (D)
  {4}         % ARG6: correct answer (1=A, 2=B, 3=C,...)
  {The last equation reads ``$0=0$'' so $x_3$ can be any 
   real number. Strictly (C) is also correct, but (D) is
   the most accurate answer.} % ARG7: solution details
  {MAH}                       % ARG8: author or source
  ...
\end{clicklist}
\end{verbatim}
\hrulefill
    \vspace*{0.2cm}
    
    The formatted output for this question (numbered Q3b-16 in the
    question bank) is displayed in \cref{fig:qexample1}. Additional
    macros are defined for the \verb+clicklist+ items
    \verb+\qitemMCthree+, \verb+\qitemMCfive+ and \verb+\qitemTF+
    corresponding to the other question types.
    
  \item \verb+clickerbeamer.{pdf,tex}+: A
    beamer~\cite{latex-beamer-2020} implementation of the
    \verb+\qitem...+ macros that instead generates a pair of slides -- one
    containing the question by itself, followed by a second showing
    both question and answer.  An example is pictured in
    \cref{fig:qexample2}.

  \item \verb+clickerslides.{tex,pdf}+: A simple non-beamer slide
    implementation. 

  \item \verb+clickerdefs.tex+: Defines several list counters and
    \LaTeX\ macros that are common to questions in all formats.

  \end{enumerate}

\item Some questions include images, plots, tables, Matlab code, or
  numerically-computed output.  All images are provided in both
  \verb+.eps+ and \verb+.pdf+ formats within the subdirectory
  \verb+figures/+.

\item All plots are generated using Matlab and the corresponding code
  can be found in the subdirectory \verb+figures/matlab/+. This permits
  any instructor to easily modify a question by changing parameters,
  plot format, etc.
\end{itemize}

The question depicted in \cref{fig:qexample1,fig:qexample2} has the
interesting feature that there are two valid responses (C,D). When this
question was actually presented in class, students were equally-split
between responses (C) and (D), which then naturally initiated a
follow-up discussion around why there were multiple correct answers and
why (D) was the ``best'' response. This is just one example of when an
instructor must be prepared to respond to poll results by spontaneously
allowing students to discuss and work through the solution for
themselves rather than just uncovering the answer slide and moving on.
Most questions in the question bank do have one unique answer, but there
are several with multiple valid responses that can present opportunities
for lively classroom debates of this sort.  Indeed, Wieman et
al.~\cite{cwsei-clicker-guide-2017} remark that ``such questions can
generate the most educationally productive discussions.''  Moreover,
clicker polls and the animated peer discussions that precede them
provide a mechanism for what mathematics educators identify as
``embodied actions to build up dynamic embodied concepts in
mathematics''~\cite{tall-2009}.

Another noteworthy feature of the question bank is that there are
several groups of closely-related questions, consisting of minor
variations on a common theme or differing levels of difficulty.  On the
one hand, this provides instructors with some variety and flexibility to
focus on specific aspects of the material.  But while it is not intended
that all such related questions would necessarily be posed together in a
single lecture, mathematical cognition researchers such as
Tall~\cite{tall-2009} have recognized that such repetitive questioning
encourages ``recognition of similarities, differences and patterns;
repetition of actions to make them routine.''

This question bank should be considered a ``work in progress'' and there
remain several gaps in coverage corresponding to topics that are part of
the standard introductory numerical analysis curriculum.  Three examples
of such gaps are Newton's method for nonlinear systems of
equations, 
matrix eigenvalue problems, 
and numerical stability and implicit methods for ODEs. 
We would also like to include more questions that illustrate concrete
applications of algorithms to problems from physics, engineering,
biology, economics or other fields, all of which we hope to include a
future release of the question bank.

\section{Mechanics of In-Class Clicker Polling}
\label{sec:polling}

\begin{figure}
  \centering
  \begin{tabular}{c@{}|c@{}|c}
    \includegraphics[width=0.27\textwidth]{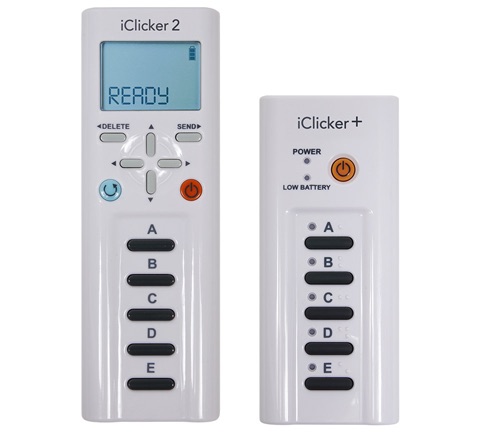}
    &
    \includegraphics[width=0.25\textwidth]{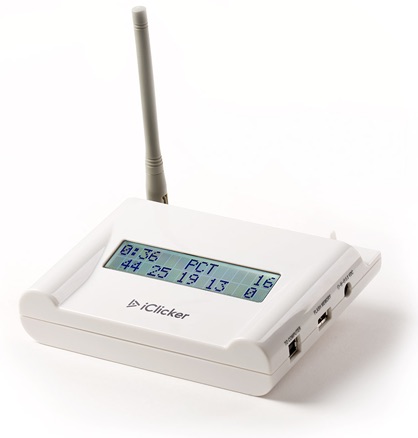}
    &
    \includegraphics[width=0.40\textwidth]{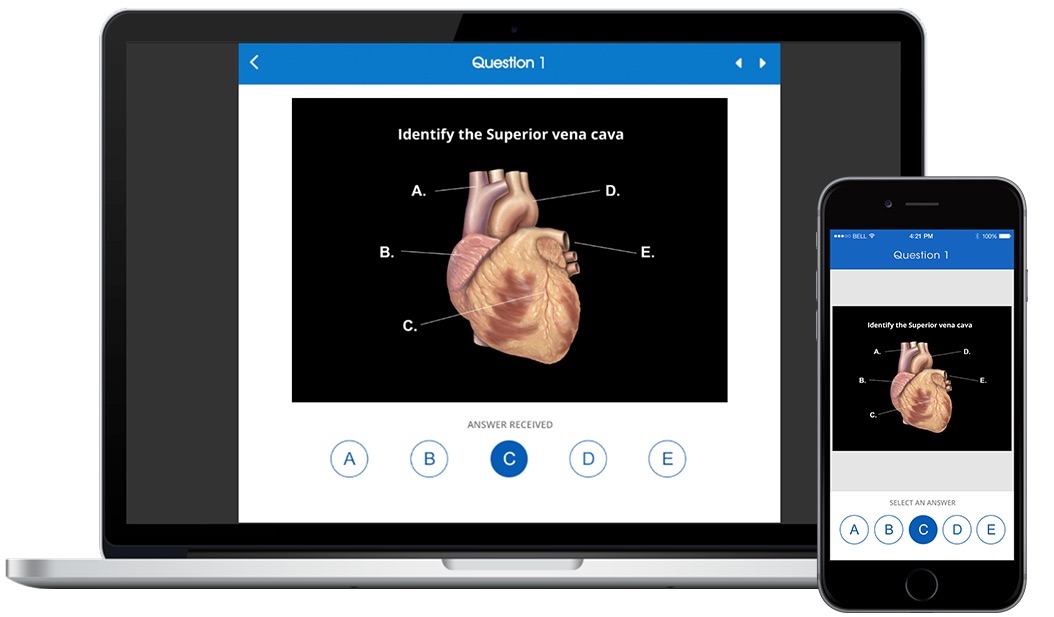}
  \end{tabular}
  \caption{(Left) Two versions of the iClicker radio-frequency remote.
    (Middle) iClicker RF base station.  (Right) Laptop and smartphone
    running the \emph{iClicker REEF} app through a Wi-Fi connection.
    Source: \url{http://www.iclicker.com}.}
  \label{fig:iclicker}
\end{figure}

For this section only, we switch to using the personal pronoun ``I''
since one of us (JMS) was the instructor for the \courseID class.  The
aim here is to provide a ``how-to guide'' of sorts that permits the
interested instructor to incorporate clicker polling into their own
lectures as quickly and easily as possible. Some of what follows relates
specifically to the iClicker system (\url{http://www.iclicker.com})
which I selected because it integrates so seamlessly with the Canvas
learning management system (\url{http://www.infrastructure.com/canvas})
used at SFU. This requires that all students purchase or borrow an
iClicker radio-frequency remote, two versions of which are pictured in
\cref{fig:iclicker}(left). Student responses are gathered by means of a
base station that is connected via USB to a computer running the
\emph{iClickerClassic} app. iClicker has an alternate web-based system
in which students use smartphones and laptops running the \emph{iClicker
  REEF} app instead, responses are transmitted through a Wi-Fi
connection, and the base station is eliminated in favour of an
internet-connected instructor app called \emph{iClicker Cloud} (shown in
\cref{fig:iclicker}(right), which also integrates with Canvas).  During
this initial trial of the question bank, I decided to use the
remote-only \emph{iClickerClassic} option so as to minimize any
difficulties managing input from multiple devices, as well as escaping
problems with intermittency in classroom Wi-Fi coverage.  However, in
future offerings of \courseID I would envisage switching to
\emph{iClicker Cloud}, not least because the most common complaint on
the student survey (see \cref{sec:survey}) was the cost of the remote
device, and students expressed a strong preference for using smartphones
to submit their responses. Indeed, it is likely that the next MACM~316
section I teach will be offered remotely due to the Covid-19 pandemic.

iClicker is only one company that markets student response systems,
and several other systems are available.  For example, Turning
Technologies manufacture their own remote devices with a corresponding
app that is also supported by Canvas. Other learning management systems
such as TopHat (\url{http://www.tophat.com}) and Acadly
(\url{http://www.acadly.com}) have their own built-in polling systems
that don't require specialized hardware but permit students to use
their smartphones or laptops.

I found that it took a few lectures to familiarize myself with the
process of clicker polling, to get used to the software, and to develop
my own routine for incorporating clicker questions into lectures. The
iClicker system makes this process relatively easy and painless, and
permits significant flexibility for customizations that suit individual
instructors. I describe below the general procedure I followed when
polling students, along with a few tips that other instructors may find
helpful:
\begin{itemize}
\item Polling is particularly easy to implement in a class where lecture
  notes are displayed using an LCD projector connected to a laptop that
  is running the iClicker app.  In \courseID, I distribute ``skeleton''
  lecture notes, typeset in \LaTeX, that provide basic definitions,
  theorems, background material and examples, and leave students to fill
  in the blanks during class. Clicker questions are easy to cut and
  paste from the \LaTeX\ question bank into my presentation slides,
  though I do not include them in the student skeleton notes 
  distributed before lectures.  The clicker questions can also be used
  in a class taught on the whiteboard or blackboard, but one has to be
  prepared to toggle back-and-forth from the LCD projector when
  displaying questions and poll results.

\item With the \emph{iClickerClassic} app running in the background, a
  small widget is displayed on top of the slide window that contains a
  play/stop button and a plot button (see \cref{fig:slide-barplot}).
  Whenever the notes are advanced to a slide containing a clicker
  question, I launch the poll from the iClicker widget which then starts
  a timer and displays a live counter of the number of responses.  I
  typically provide students between 45--60 seconds to select a response
  and I also give a 10-second warning before closing each poll.  It is
  possible to configure the app to use a count-down timer instead so
  that students can see precisely how much time is left. On completing
  a poll, clicking the plot button on the widget displays a histogram
  of student responses (see \cref{fig:slide-barplot}).%

  \begin{figure}
    \centering
    \fbox{\includegraphics[width=0.9\textwidth]{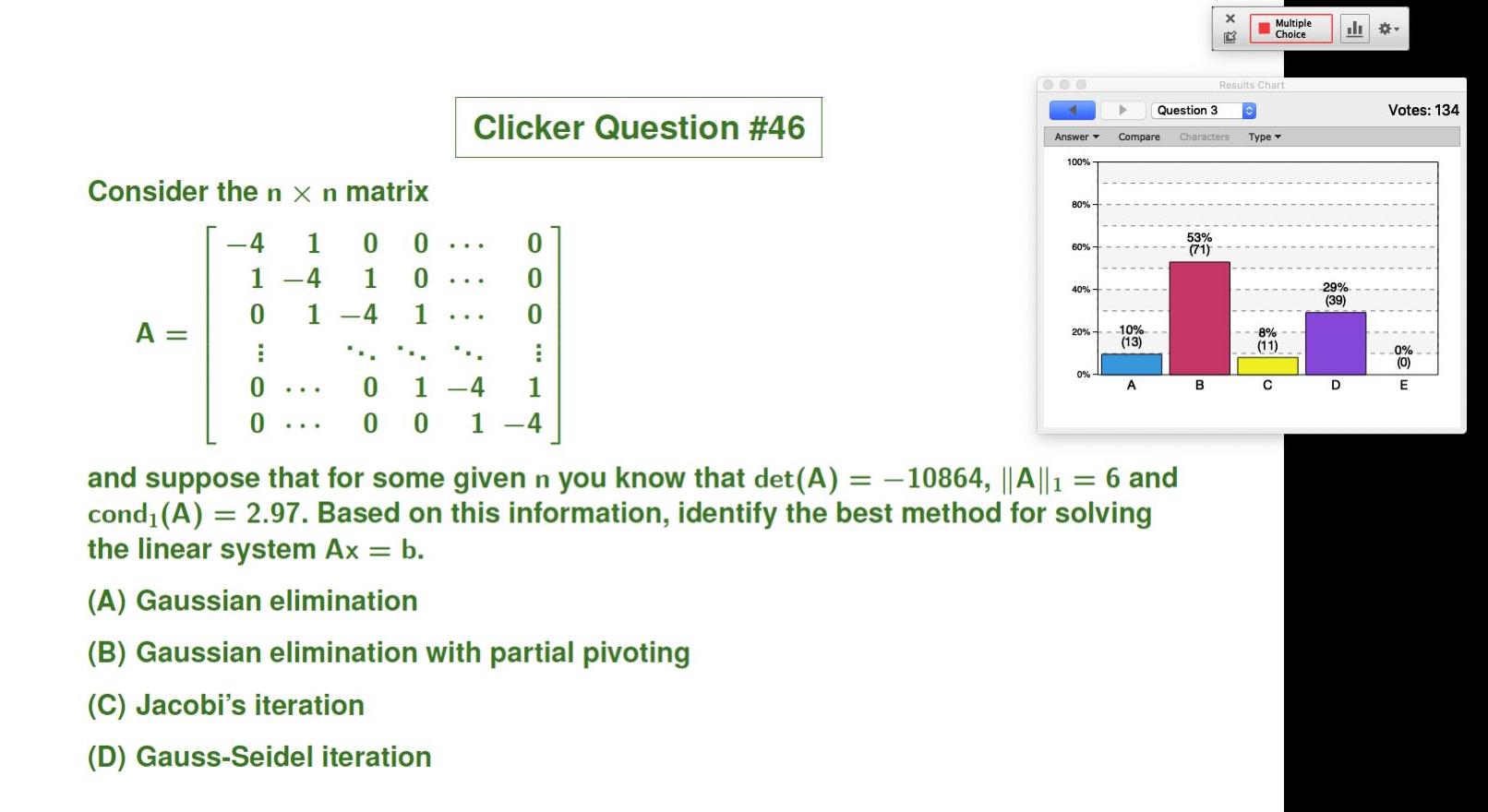}}
    \caption{Lecture slide displaying a clicker question, overlaid by
      the \emph{iClicker Classic} widget (top right) and a bar plot
      summarizing student responses (middle right).}
    \label{fig:slide-barplot}
  \end{figure}

\item I normally pose between 3 to 5 clicker questions during each
  50-minute lecture, with the number depending on the material being
  taught.  The questions are grouped in clusters of 2-3 to minimize the
  overhead required when asking students to get their clickers ready,
  changing between applications, and ``switching gears'' in my
  delivery. Working through each question cluster typically requires at
  least 5 minutes, which is not only a welcome break for students but
  also helps me avoid my own natural tendency to speed up and
  ``steam-roll'' through the lecture material.

\item I imposed no set pattern on the timing of questions in any given
  lecture, variably using them as initial warm-up problems, or to check
  in on student understanding after teaching a new concept, or as a
  final wrap-up, or simply to introduce a break.  The key is to be
  flexible.

\item I encourage students to discuss questions with their neighbours
  and I occasionally walk around the room to listen in on conversations
  and gauge what difficulties students might be having.  Quieting
  students down after a particularly vigorous discussion may sometimes
  take a few extra moments, but I believe that this is time well spent
  and not at all disruptive.
 
\item Upon revealing the histogram of poll results, if the majority of
  students chose the correct answer (at least 75\%) then I normally
  advance directly to the solution slide, highlight an important point
  or two, and move on. Otherwise, time permitting, I devote some
  extra time to a follow-up discussion or work through some
  solution details to ensure that students who answered incorrectly
  understand the source of their error.

\item If a question receives an unusually high proportion of incorrect
  responses then I make a special note of it, recognizing that such
  questions are typically either:
  \begin{itemize}
  \item overly difficult, ill-posed or confusingly worded: such
    questions were either be modified or removed from the question bank.

  \item challenging or ``tricky'': which are often good topics for extra
    discussion. I flag these equations as potential test material and/or
    devise supplementary examples that explore deeper aspects of the
    material for future classes.
  \end{itemize}
\end{itemize}
It is important to recognize that any extra class time spent on clicker
polling and discussion has a significant effect on the material that can
be covered. Wieman et al.~\cite{cwsei-clicker-guide-2017} observe that
``most instructors do end up deciding to cover slightly ($\sim$10\%)
less material in their courses after they have started using clickers
effectively,'' and my own experience is consistent with this. However,
this is a small price to pay considering the substantial benefits in
terms of student satisfaction and improved classroom learning
environment.

\mysubsection{Unexpected Switch to On-line Instruction} 

The Jan-Apr 2020 semester on which this paper is based was an
exceedingly unusual one because it coincided with the outbreak of the
Covid-19 pandemic. Classes at SFU ran as usual until mid-March
(essentially mid-semester) at which time the university decided to shift
all classes to an on-line mode of delivery to comply with provincial
physical distancing guidelines. Because iClicker remotes function only
in a classroom setting with an RF base station, I decided at that point
to stop collecting clicker response data and to determine students'
participation grades based only on their responses from the first
half-semester of clicker questions.

Like many other universities, SFU has responded to the Covid-19 pandemic
by moving almost exclusively to an on-line mode of teaching for the next
academic year.  This means that clicker remotes cannot be used since
they operate by communicating radio-frequency signals to a base station.
Therefore, at least in the near future, clicker polling in \courseID
will have to be done via Wi-Fi connection using a system like
\emph{iClicker Cloud/REEF}.

Although the mid-semester interruption of this clicker experiment was
very disheartening, it did uncover another convenient but unforeseen
application of the clicker question bank. Upon the switch to on-line
instruction, I simultaneously transformed the weekly in-person quizzes
into an on-line multiple-choice format for which it seemed natural to
re-purpose clicker questions. Based on this experience, I can confirm
that the clicker bank is also an excellent source of multiple-choice and
true-false questions for use on quizzes or tests.

\section{Student Satisfaction Survey Results}
\label{sec:survey}

A student satisfaction survey was held at the mid-point of semester,
which coincidentally was only two days before SFU switched all classes
to an on-line mode of instruction in response to the Covid-19
pandemic. It was unfortunate that the in-class clicker polling had to be
terminated at that point owing to the reliance on iClicker
remotes. However, because the survey was completed before the switch, we
are confident that the results are unaffected by any of the disruptions
that ensued. 

A summary of all survey questions and response data is given in
\cref{app:survey-data}.  The survey consisted of 7 multiple-choice
questions (Q1--Q7), another set of 10 statements rated on a Likert scale
(Q8), and one free-response comment box (Q9). Several questions were
modelled after similar surveys of clicker-based calculus
classes~\cite{bode-etal-2009, lucas-2009, menz-jungic-wiebe-2009}, and
the Likert-scale question Q8 was inspired by the human development class
in \cite{beckert-fauth-olsen-2009}. A total of 137 responses were
received out of 199 registered students. The corresponding response rate
of 69\%\ is relatively high for this sort of survey and so we are
confident that the results are representative of the entire class. The
survey was administered as a Canvas Quiz, which allowed responses to be
submitted anonymously and the data easily downloaded.

Rather than describing all survey responses in detail, we will highlight
what we view as the most important results and refer the interested
reader to \cref{app:survey-data} for more specifics.
\begin{itemize}
\item Questions Q2 and Q3 suggest that more than 80\%\ of students
  thought the clicker questions helped them to improve their
  understanding of the course material.

\item Questions Q4 through Q7 indicate that the majority of students
  were satisfied with the level of difficulty of the problems, the
  method of classroom delivery, and the grading scheme based on
  participation only.  Some instructors prefer to assign a clicker grade
  that gives credit for a mix of participation and
  correctness~\cite{cwsei-clicker-guide-2017} and this is easy to set up
  within the iClicker system.  But our experience survey suggests that
  students derive sufficient benefit without pressuring them to submit
  correct responses.

\item Question Q8 is an especially strong indicator of the high level of
  satisfaction and positive impacts on students' learning
  experience. The first three statements (A,B,C) all received more than
  80\%\ positive responses (either ``agree'' or ``strongly agree'') on
  the Likert scale, which reflects two important aspects of clicker
  polling that are also recognized in other studies:
  \begin{itemize}
  \item Students have a strong preference for the anonym\-i\-ty of a
    clicker poll over answering questions individually, most likely due
    to shyness or fear of embarrassment in a large class.
  \item Clickers provide a positive break from the rest of lecture.
  \end{itemize}
  Of the remaining statements (D to J) all but H received at least 50\%\
  positive responses and at most 20\%\ negative, which reveals the
  beneficial contributions of clicker polling to the classroom learning
  environment.

\item One result from question Q8 that deserves special mention concerns
  statement H (``I feel more connected to my classmates when we use
  iClickers'') which had a more mixed response. This exposes a distinct
  opportunity for us to involve students more actively and
  collaboratively with each other by incorporating a \emph{peer
    instruction} component into the clicker polling process (sometimes
  also referred to as ``think--pair--share''). In this approach,
  students are provided extra time after the initial clicker poll (but
  before the solution is displayed) to discuss the poll results with
  each other and then a second poll is taken to determine any changes in
  student opinion. Peer instruction is a well-established teaching
  practice~\cite{crouch-mazur-2001, lucas-2009, menz-jungic-wiebe-2009,
    miller-etal-2006, pilzer-2001, cwsei-clicker-guide-2017} that has
  proven effective in engaging students more actively in the learning
  process, improving their critical thinking and communication skills,
  and building a more collegial intellectual atmosphere. This is
  something that obviously consumes additional class time, but can be
  introduced gradually and is not necessary to do for every clicker
  question.
\end{itemize}

\mysubsection{Free-form Comments}

The final survey question asked students to provide free-form comments
in response to the prompt: ``Do you have any other comments or
suggestions related to the iClicker questions?''.  The feedback we
received was without doubt the most encouraging indicator of student
satisfaction with the clicker experiment.  

First of all, let us summarize the critical (but mostly constructive)
comments we received which can be distilled down to essentially four
points:
\begin{itemize}
\item \emph{iClicker remotes are too expensive:} The cost of roughly
  \$50 for a remote is a significant expense, although not unreasonable
  when one considers that nearly 80\%\ of the MACM~316 students needed
  them in at least one other class (Q1). It is not surprising that
  students would prefer to use their smartphones, but while the
  \emph{iClicker REEF} app can be downloaded freely it does require an
  annual subscription fee of \$25. One possible concern about
  encouraging students to use their smartphones in class is that they
  may be more easily distracted by alerts from social media and other
  extraneous apps on their devices.

\item \emph{Can you post clicker questions and answers after class?:} In
  our opinion, the clicker component of this course is a major benefit
  of attending lectures in person. We have also found that certain
  clicker questions make excellent test or exam questions, which is why
  we prefered to not make the questions available to students. Wieman et
  al.~\cite{cwsei-clicker-guide-2017} recognize that there are varying
  opinions about whether or not to provide clicker questions and
  solutions after class, and conclude that there is ``no data that
  indicates one way or another.''

\item \emph{Polling time was insufficient, or the question text was too
    long:} This is a valid complaint and we have already made attempts
  to shorten the text in some of the wordier questions. We also plan in
  future to devise more short questions, in particular those of the
  true-false type.

\item \emph{Questions were spread unevenly between lectures:} Although
  our original plan was to evenly distribute 3--5 questions in every
  lecture, we failed to account sufficiently for the unpredictable
  delays involved with conducting polls and discussing the
  results. Based on our experience, we plan to modify the lecture notes
  and adjust the distribution of questions to address this concern.
\end{itemize}
Outside of these few critical comments, we were gratified by the many
more remarks of a positive nature that highlighted students' high degree
of satisfaction with their learning experience.  We have listed a few
selected comments below, which are far from an exhaustive list but
hopefully convey some sense of the range of potential benefits that
clicker polling can have on the learning environment:
\begin{itemize}
\item ``I think iClickers serve as a useful addition to a class
  setting. It encourages active participation and encourages students to
  engage with the material. In this sense, it seems to work well,
  especially in a large class setting.''

\item ``It helps to keep my attention in class when we're
  switching back and forth (helps with short attention span).''

\item ``I do not feel the pressure of getting a wrong answer in front of
  a 150 person crowd. I also have more time to think about an answer
  rather than a sharp student immediately answering so I get more out of
  the question.'' 

\item ``At the beginning of the semester I was slightly annoyed with
  having to buy an iClicker. It was used quite effectively though, and I
  feel that it was worth the cost overall. It gave me a little extra
  push to make it to class, without the panic that I get from graded
  quizzes. I definitely prefer it to the alternatives like TopHat or
  Canvas quizzes, which are difficult to connect to in some
  classrooms.''

\item ``I usually enjoy classes with iClickers more because I usually
  dislike raising my hand so iClickers grant a anonymity that I enjoy.''

\item ``The depth and quality of questions asked made up for the cost
  and then some. It also helped knowing others in class via 
  discussions, and the a\-no\-nym\-i\-ty was helpful to know sometimes I
  am not the only one confused.''

\item ``The participation is a very positive way of doing the questions,
  rather than pressuring students to get them right. Learning is
  encouraged, rather than only focusing about getting the correct
  answer.''

\item ``I love your usage of iClickers, it acts like a little break in
  between class that helps us keep focused more before/after we do
  them. They also let us learn at our own pace, getting an answer wrong
  means I need to work on that certain material again.''

\item ``The questions gave me immediate feedback on how I was doing in
  the class and what I need more practice on.''

\item ``I particularly like tricky questions, edge case questions, or
  questions with an unintuitive answer. I feel like these questions
  really help to solidify my understanding of the material.'' 
\end{itemize}

\mysubsection{Anecdotal Observations} 

We make no claim that clicker polling has had any direct connection with
improvements in students' grades or other quantifiable learning
outcomes.  Some previous studies have attempted to do
so~\cite{crouch-mazur-2001, lucas-2009, miller-etal-2006, pilzer-2001,
  roth-2012} and although the results suggest there are some positive
correlations with student performance, this effect tends to be
relatively small. Our focus in this paper is rather on the significant
impact that devoting class time to clickers can have on the classroom
atmosphere and students' learning experience.  Outside of the student
survey described above, we have observations from the instructor's
perspective of a ``before and after'' nature (comparing the most recent
semester with two other recent offerings of the same course) that
provide anecdotal support for the positive effects of clickers in the
classroom:
\begin{itemize}
\item \emph{Increased attendance:} The percentage of enrolled students
  attending lectures (when there was no in-class quiz) was between
  70--85\%\ and normally lay closer to the high end of this range.  This
  is in comparison with a 50\%\ attendance estimated from other recent
  offerings of the course.  It should be no surprise whatsoever that
  assigning a portion of the grade that requires attendance in class
  should increase attendance. Rather, we want to highlight the virtual
  absence of any negative feedback from the survey regarding the clicker
  participation grade, nor was there any suggestion that this was viewed
  as some form of ``coercion''.

\item \emph{Student questions before/after lectures:} Compared with
  previous semesters, there was a marked increase in the number of
  students asking questions during class time (outside that devoted to
  clicker polling). But what was particularly striking was the long
  line-up of students waiting to ask questions after class, which
  often required an additional 20-30 minutes. It is unusual to
  see such enthusiasm from a large number of students, and we can
  only hypothesize that perhaps the clicker polling gave students a
  sense that their instructor is more approachable and cares about their
  learning.

\item \emph{On-line discussions:} A discussion board has always been
  provided as part of the Canvas container for this class, and there was
  more than a 100\%\ increase in student postings \emph{before} the move
  to on-line instruction at mid-semester.
\end{itemize}
When combined with the survey results, these observations suggest that
engaging students more actively in the classroom through the use of
clicker polling can have certain beneficial side-effects in terms of not
only improving the students' sense of engagement but also enhancing the
atmosphere of collegiality both inside and outside of the classroom.

\section{Closing remarks}
\label{sec:remarks}

The use of clicker polling in university mathematics class\-es is
commonplace and while many clicker question resources are available for
introductory courses, there is next to nothing available for more
advanced courses such as numerical analysis. We describe an open-access
clicker question bank that we have developed to fill this gap, and which 
covers a significant portion of the introductory numerical analysis
curriculum. Based on the first semester in which the questions were
implemented in class, results from a student survey suggest that
clickers had a significant positive impact on the student learning
experience.


We would like to close with a quote taken from Wieman et
al.~\cite{cwsei-clicker-guide-2017}: 
\begin{quote}
  \itshape
  ``We have found the effective use of clicker questions and discussion
  can have a transformative impact on both teachers and students,
  particularly in large classes. Students end up being far more actively
  engaged in the material and they learn more, and both students and
  instructors find the course much more rewarding as a result.''
\end{quote}
This has certainly been our experience and we strongly encourage readers
to consider using clicker polling in their own classroom. Furthermore,
with so many universities around the world recently transitioning to
remote class delivery in response to the Covid-19 pandemic, clickers
present a special opportunity for instructors to engage their students
actively in an on-line lecture setting. It is our sincere hope that the
open-access clicker question bank described here will help to reduce the
``cost of entry'' for instructors of numerical analysis who have an
interest in experimenting with student response systems.
  
\appendix
\section{Student Survey Data}
\label{app:survey-data}

For the first 7 questions (Q1--Q7) the responses are depicted as a
bar chart showing the percentage of all 137 completed surveys, and the
number of students who selected each response is also shown in parentheses. 

\newcommand{\myrule}{\rule{0cm}{0.4cm}}
\newcommand{\myrulehalf}{\rule{0cm}{0.4cm}}

\newcommand{\clickqtwo}[6]{%
  \item #1
  \begin{center}
    \footnotesize
    \begin{minipage}{0.50\linewidth}
      \begin{flushright}
        #2 (#3)\\[1.1cm]
        #4 (#5)\\[0.2cm]
        \mbox{}
      \end{flushright}
    \end{minipage}
    \begin{minipage}{0.48\linewidth}
      \includegraphics[trim=100 100 100 100,clip,width=\textwidth]{#6}
    \end{minipage}
  \end{center}}

\newcommand{\clickqtwoMOD}[6]{%
  \item #1
  \begin{center}
    \scriptsize
    \begin{minipage}{0.53\linewidth}
      \begin{flushright}
        #2 (#3)\\[0.9cm]
        #4 (#5)\\[0.2cm]
        \mbox{}
      \end{flushright}
    \end{minipage}
    \begin{minipage}{0.45\linewidth}
      \includegraphics[trim=105 100 110 100,clip,width=\textwidth]{#6}
    \end{minipage}
  \end{center}}

\newcommand{\clickqthree}[8]{%
  \item #1
  \begin{center}
    \footnotesize
    \begin{minipage}{0.50\linewidth}
      \vspace*{-0.5cm}
      \begin{flushright}
        #2 (#3)\\[0.3cm]
        #4 (#5)\\[0.3cm]
        #6 (#7)
      \end{flushright}
    \end{minipage}
    \begin{minipage}{0.48\linewidth}
      \includegraphics[trim=100 100 100 100,clip,width=\textwidth]{#8}
    \end{minipage}
  \end{center}}

\newcommand{\clickqthreeMOD}[8]{%
  \item #1
  \begin{center}
    \scriptsize
    \begin{minipage}{0.53\linewidth}
      \vspace*{-0.3cm}
      \begin{flushright}
        #2 (#3)\\[0.4cm]
        #4 (#5)\\[0.4cm]
        #6 (#7)
      \end{flushright}
    \end{minipage}
    \begin{minipage}{0.45\linewidth}
      \includegraphics[trim=105 100 110 100,clip,width=\textwidth]{#8}
    \end{minipage}
  \end{center}}

\def\clickqfour#1#2#3#4#5#6#7#8#9{%
  \def\tempq{#1}%
  \def\temprespone{#2}%
  \def\tempnumone{#3}%
  \def\tempresptwo{#4}%
  \def\tempnumtwo{#5}%
  \def\temprespthree{#6}%
  \def\tempnumthree{#7}%
  \def\temprespfour{#8}%
  \def\tempnumfour{#9}%
  \clickqfourcontd}
\newcommand{\clickqfourcontd}[1]{%
\item \tempq
  \begin{center}
    \footnotesize
    \begin{minipage}{0.50\linewidth}
      \begin{flushright}
        \temprespone\ (\tempnumone)\\[0.2cm]
        \tempresptwo\ (\tempnumtwo)\\[0.2cm]
        \temprespthree\ (\tempnumthree)\\[0.2cm]
        \temprespfour\ (\tempnumfour)
      \end{flushright}
    \end{minipage}
    \begin{minipage}{0.48\linewidth}
      \includegraphics[trim=100 100 100 100,clip,width=\textwidth]{#1}
    \end{minipage}
  \end{center}}

\def\clickqfourMOD#1#2#3#4#5#6#7#8#9{%
  \def\tempq{#1}%
  \def\temprespone{#2}%
  \def\tempnumone{#3}%
  \def\tempresptwo{#4}%
  \def\tempnumtwo{#5}%
  \def\temprespthree{#6}%
  \def\tempnumthree{#7}%
  \def\temprespfour{#8}%
  \def\tempnumfour{#9}%
  \clickqfourcontdMOD}
\newcommand{\clickqfourcontdMOD}[1]{%
\item \tempq
  \begin{center}
    \scriptsize
    \begin{minipage}{0.53\linewidth}
      \begin{flushright}
        \temprespone\ (\tempnumone)\\[0.1cm]
        \tempresptwo\ (\tempnumtwo)\\[0.1cm]
        \temprespthree\ (\tempnumthree)\\[0.1cm]
        \temprespfour\ (\tempnumfour)
      \end{flushright}
    \end{minipage}
    \begin{minipage}{0.45\linewidth}
      \includegraphics[trim=105 100 110 100,clip,width=\textwidth]{#1}
    \end{minipage}
  \end{center}}

\newcommand{\myvspace}{\vspace*{0.311cm}}
\newcommand{\myvspaceMOD}{\vspace*{0.25cm}}

\begin{enumerate}[label=Q\arabic*.]
  \clickqthreeMOD{In how many other classes have you used iClickers before
    this semester?}{None\myrule}{30}{One class\myrule}{24}{More than one
    class\myrule}{83}{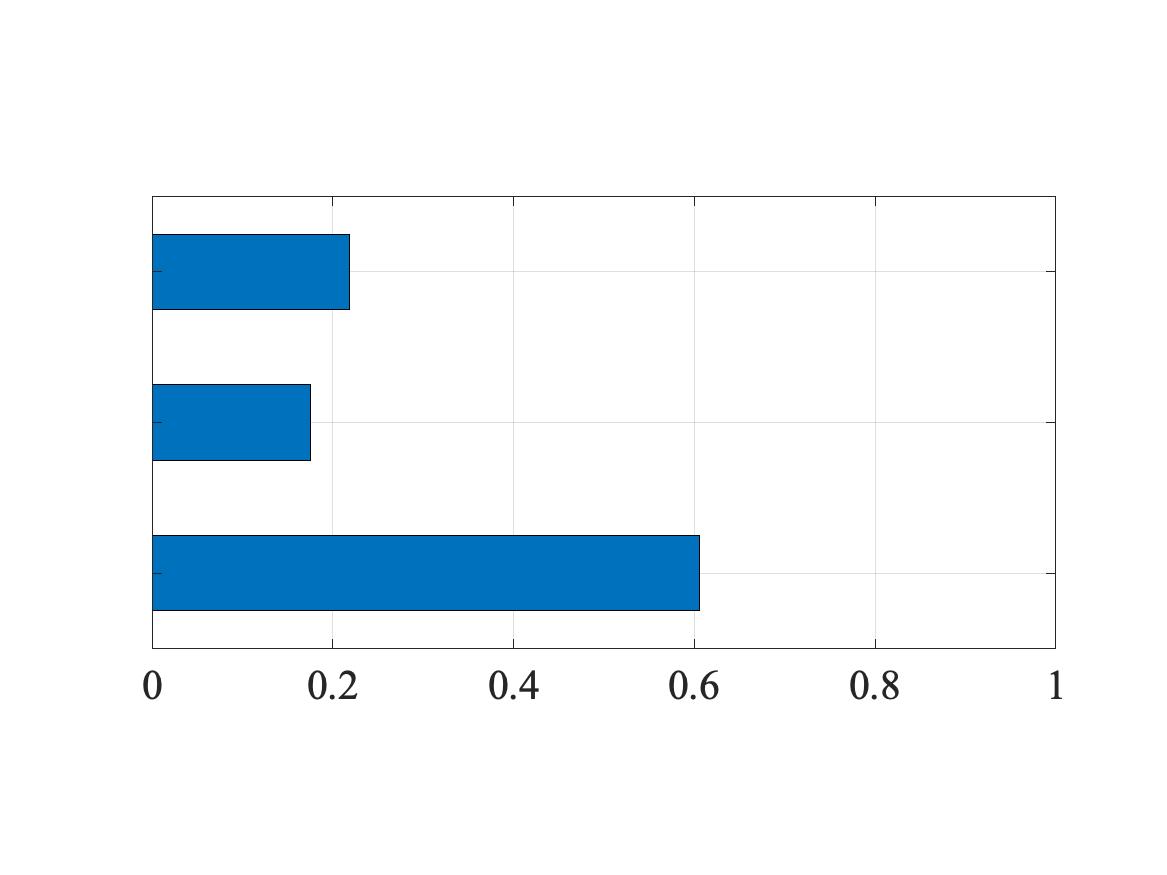}
  
  \clickqtwoMOD{Overall, do you find that the iClicker questions have
    helped you to better understand the material presented in
    lectures?}{Yes\myrule}{116}{No\myrule}{21}{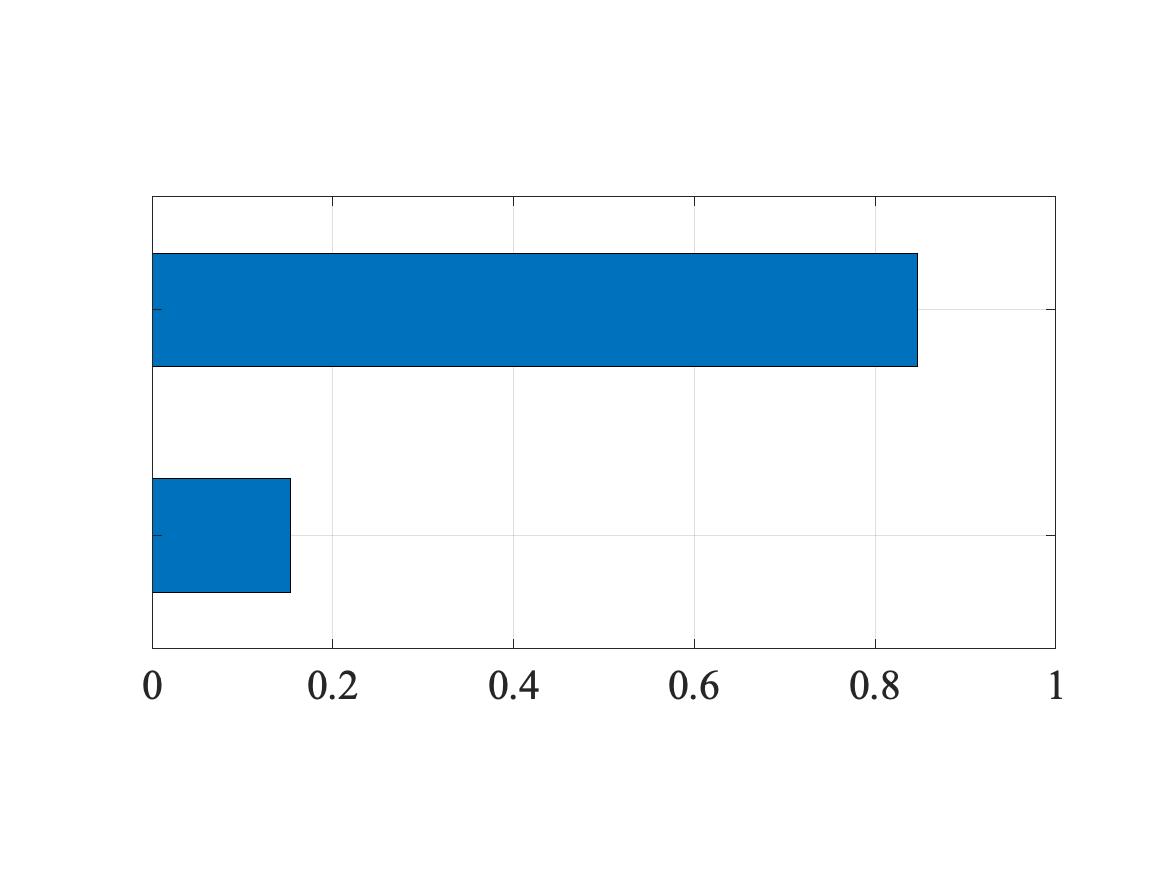}

  \clickqthreeMOD{Pick the statement that best describes your experience
    with the iClicker questions.}{They are good practice to make sure I
    understand concepts from lectures}{112}{I often get the
    wrong answer and am left confused, even after the correct answer is
    discussed}{16}{I don't get anything out of it and would
    rather focus on plowing through the lecture
    material}{9}{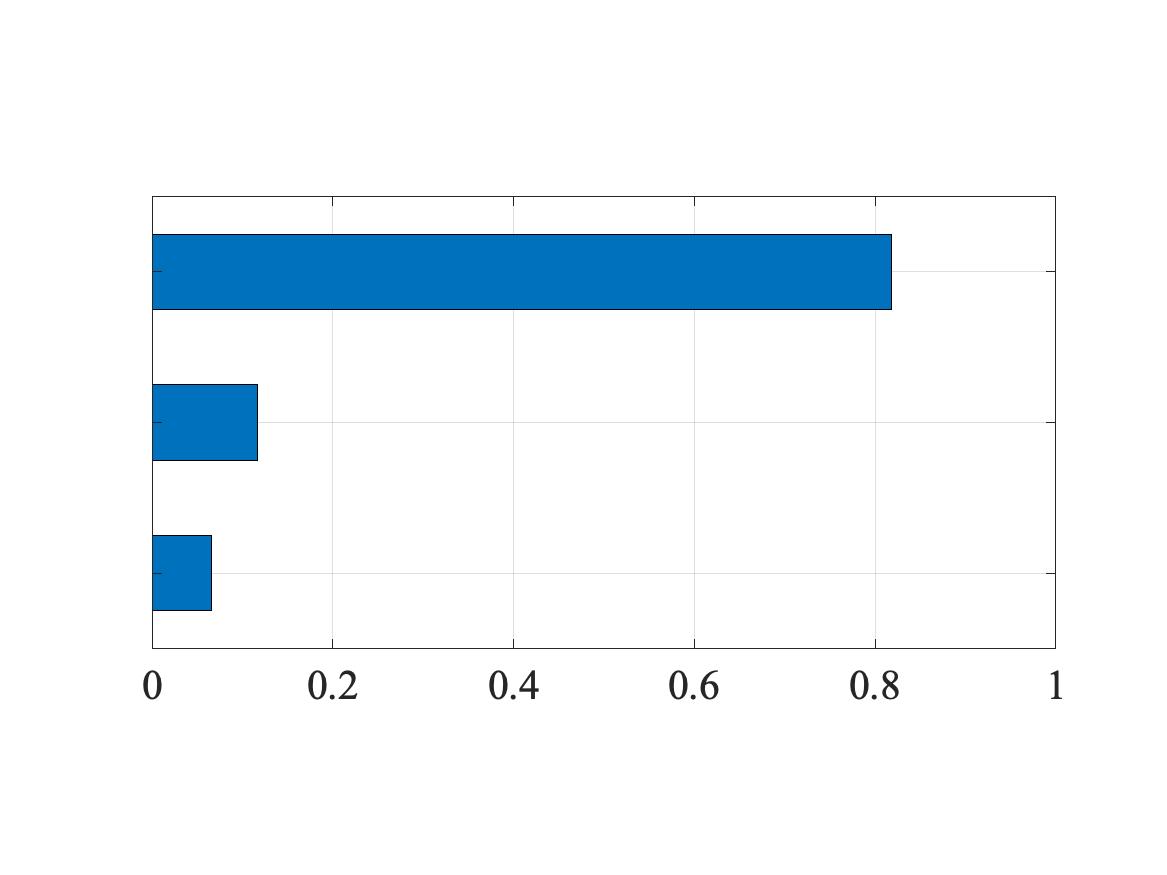}

  \clickqthreeMOD{Can you comment on the amount of time permitted to
    answer each question?}{Too much time -- I was often waiting too long
    after keying in my answer}{3}{The time allowed was just
    right}{81}{Too little time -- I felt rushed and often had
    to key in a random response at the last
    minute}{53}{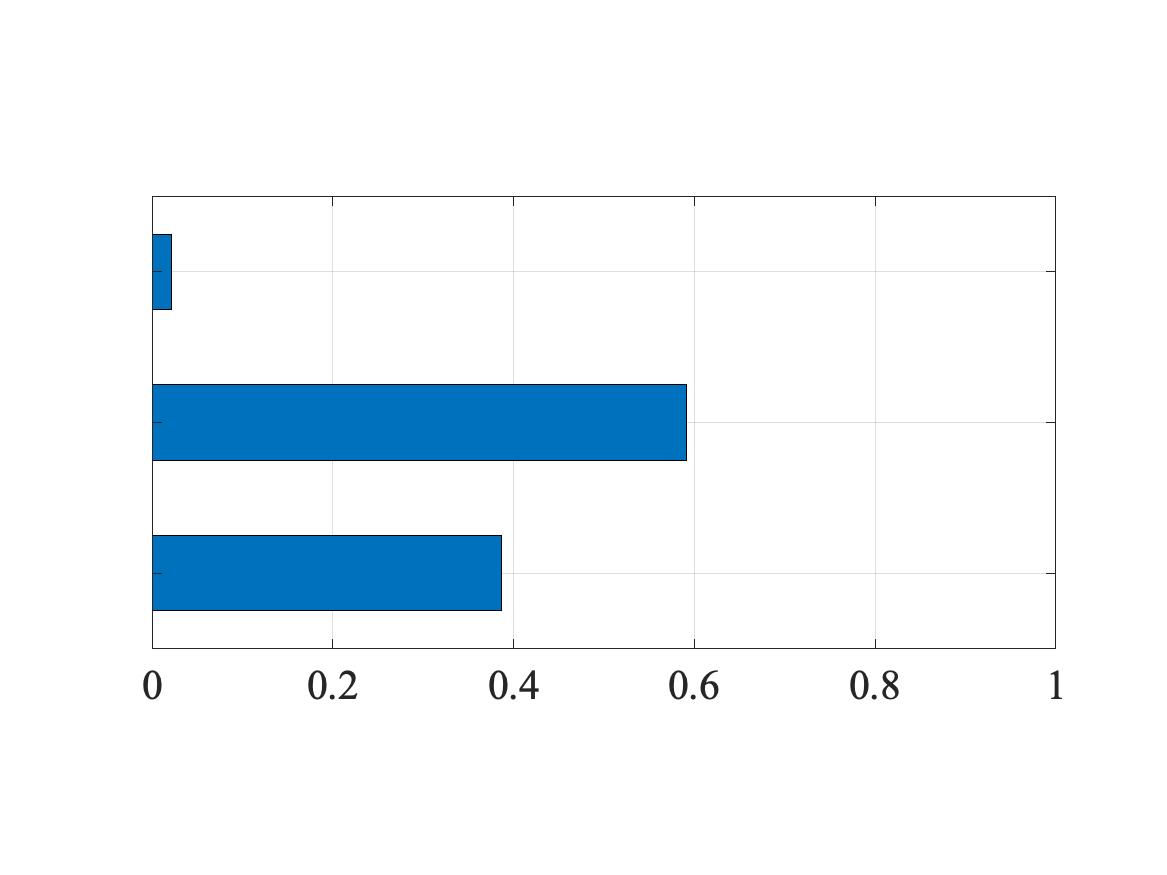}

  \clickqthreeMOD{Can you comment on the average level of difficulty of
    the questions?}{Most were too difficult\myrulehalf}{13}{Just right --
    there was a good mix of easy/hard questions}{121}{Most were too
    easy}{3}{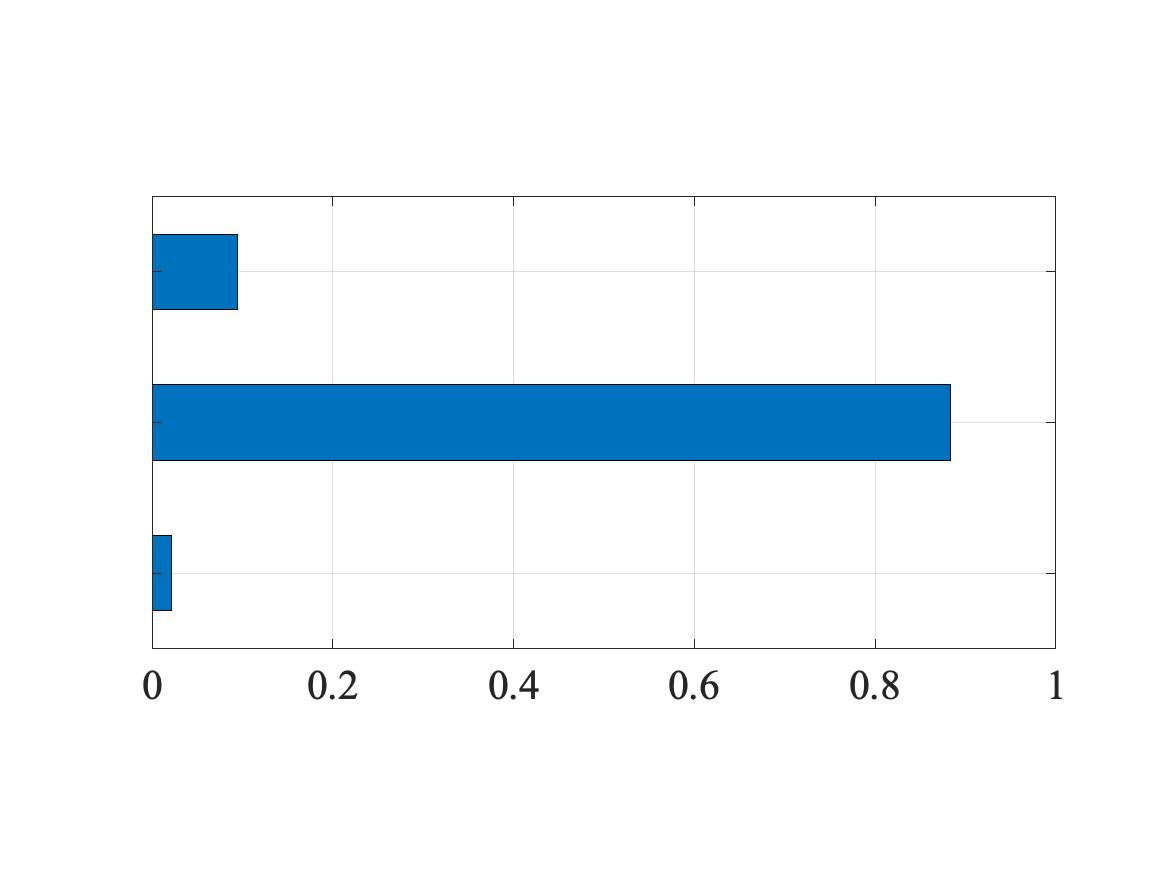}

  \clickqfourMOD{Do you have any preference regarding how the iClicker
    grade should be computed?}{I like the current formula: participation
    grade only, no marks for correctness}{117}{Assign no marks for
    participation, and give credit for correct answers only}{4}{Assign a
    combined grade that credits both participation and correct
    responses}{7}{No marks should be assigned to the iClicker
    questions}{9}{q6}

  \clickqthreeMOD{Assuming that some marks are assigned to completing the
    iClicker questions, what is your opinion of the portion 3\%\ 
    allocated to your final grade?}{3\%\ is just right\myrule}{79}{It
    should be worth more than 3\%\myrule}{46}{It should be worth less
    than 3\%\myrule}{12}{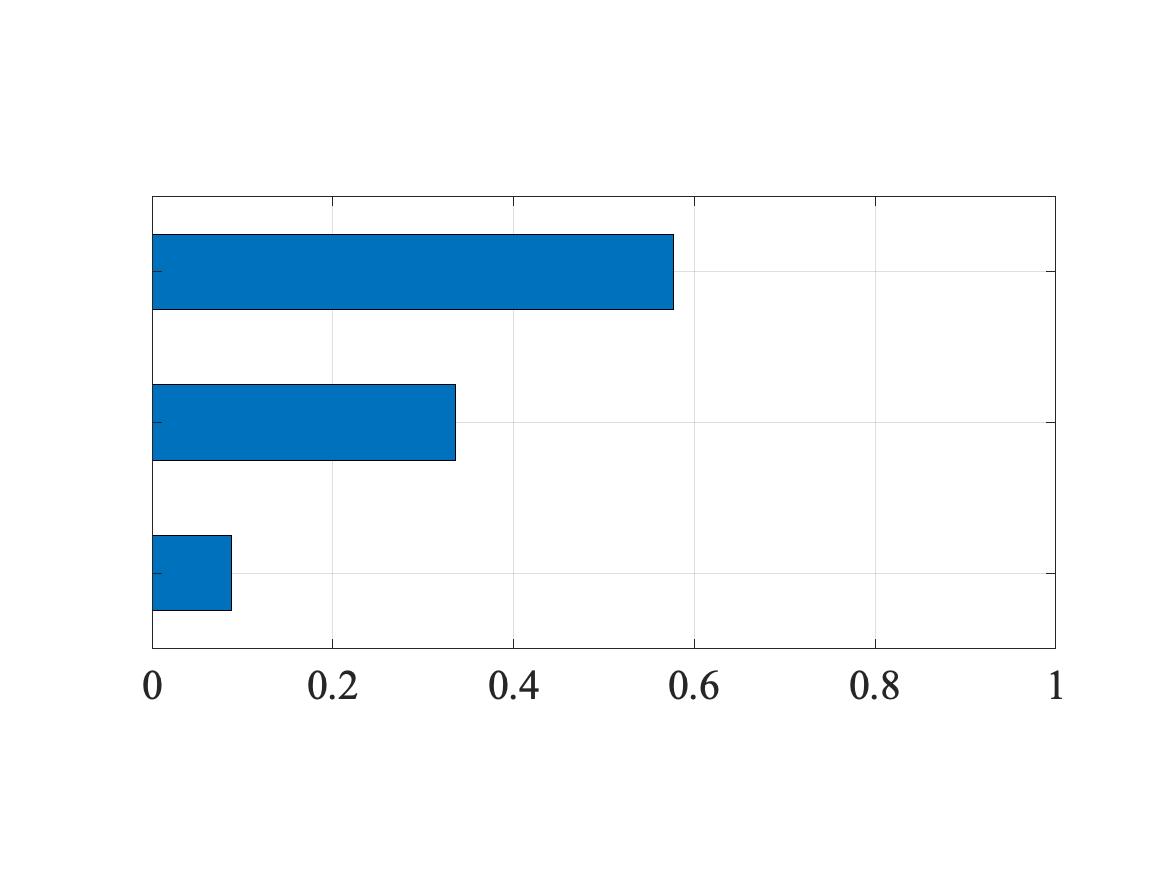}

\item Rate your level of agreement with statements A-J on the
  following scale:
  \vspace*{0.2cm}
  \begin{center}
    -- Strongly Disagree, Disagree, Neutral, 
    Agree, Strongly Agree --
  \end{center}
  \vspace*{0.2cm}

  \begin{enumerate}[label=\Alph*.]
  \item \mbox{[raise-hand]} When a question is asked, I like the
    iClicker response option better than asking students to raise their
    hands to indicate their answer.

  \item \mbox{[volunteer]} When a question is asked, I like the iClicker
    response option better than having an open invitation for any
    student to volunteer an answer.

  \item \mbox{[positive-break]} I think iClickers provide a positive
    break from lectures.
    
  \item \mbox{[engaged]} I feel more involved and engaged in class when
    we use iClickers.
    
  \item \mbox{[participation]} I come to class more often because
    iClickers are used to track participation.

  \item \mbox{[retain-info]} I retain more information from lecture when
    iClickers are used.

  \item \mbox{[distraction]} I think iClickers distract from the learning
    process.  {\LARGE\color{red}$\pmb{\ast}$}

  \item \mbox{[connection]} I feel more connected to my classmates when
    we use iClickers.

  \item \mbox{[insights]} iClickers help me to discover more insights
    into the course material.
    
  \item \mbox{[attention]} It is easier to pay attention in lectures when
    iClickers are used.
  \end{enumerate}
  Note that statement G (marked {\LARGE\color{red}$\pmb{\ast}$}) is the
  only one where a response of ``strongly agree'' has a \emph{negative}
  connotation, and so this bar should be interpreted in the opposite
  manner to the others.  This was done to verify that students were
  indeed carefully reading and applying the Likert scale to each
  statement.

  \vspace*{0.2cm}
  \begin{center}
    \footnotesize
    \begin{minipage}{0.18\linewidth}
      \vspace*{-0.2cm}
      \begin{flushright}
        [raise-hand] \\    \myvspaceMOD
        [volunteer] \\     \myvspaceMOD
        [positive-break] \\\myvspaceMOD
        [engaged] \\       \myvspaceMOD
        [participation] \\ \myvspaceMOD
        [retain-info] \\   \myvspaceMOD
        {\color{red}[distraction]}\\\myvspaceMOD
        [connection] \\    \myvspaceMOD
        [insights] \\      \myvspaceMOD
        [attention]
      \end{flushright}
    \end{minipage}
    \begin{minipage}{0.8\linewidth}
      \includegraphics[trim=100 10 80 60,clip,width=\textwidth]{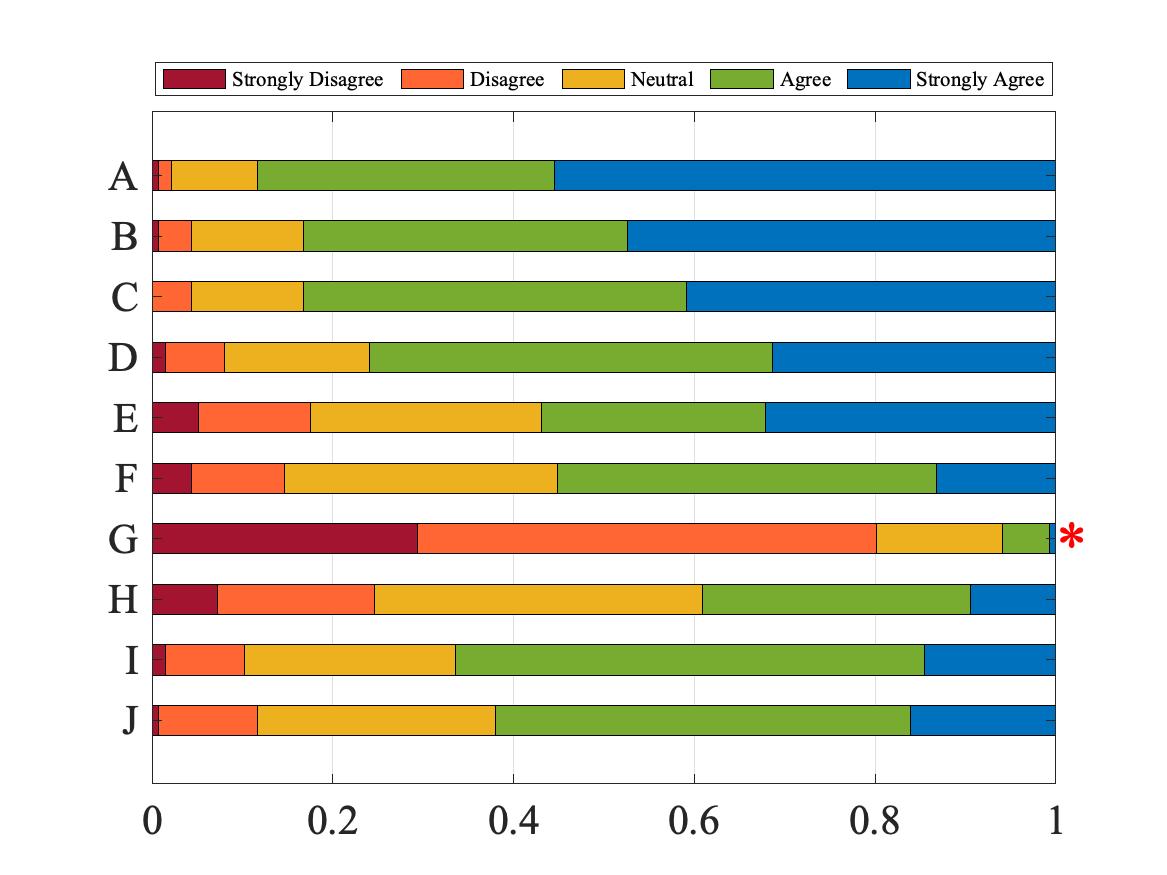}
    \end{minipage}
  \end{center}
\end{enumerate}




\bibliographystyle{siamplain}

\providecommand{\noopsort}[1]{}

\end{document}